\documentclass[%
reprint,aps,amsmath,amssymb,prb,superscriptaddress,longbibliography%
]{revtex4-2}
\usepackage{dcolumn} 
\usepackage{graphicx} 
\usepackage{graphicx,subcaption}
\usepackage[version=3]{mhchem}
\usepackage{color,soul}
\usepackage{multirow}
\usepackage{braket}
\usepackage{xcolor}
\usepackage{bm}
\usepackage{comment}
\usepackage{float}

\newcommand{\bracket}[3]{\left< #1\left|#2\right|#3 \right>} 
\newcommand{\hbm}[1]{\hat{\bm{#1}}}
\newcommand{\grad}[1]{\nabla_{\bm{#1}}}
\newcommand{\pp}[2]{\frac{\partial{#1}}{\partial{#2}}}

\newcommand{\bGamma}{{\bm \Gamma}}

\newcommand{\hH}{\hat H}

\newcommand{\bP}{{\bm P}}
\newcommand{\bX}{{\bm X}}

\newcommand{\bd}{{\bm d}}

\newcommand{\bK}{{\bm K}}
\newcommand{\bI}{{\bm I}}

\begin{document}

\title{A Basis-Free Phase Space Electronic Hamiltonian That Recovers Beyond Born-Oppenheimer Electronic Momentum and Current Density}

\author{Zhen Tao}
\email{zhen.tao@princeton.edu}
\affiliation{Department of Chemistry, Princeton University, Princeton, NJ USA}
\author{Tian Qiu}
\email{tq4143@princeton.edu}
\affiliation{Department of Chemistry, Princeton University, Princeton, NJ USA}
\author{Xuezhi Bian}
\affiliation{Department of Chemistry, Princeton University, Princeton, NJ USA}
\author{Joseph E. Subotnik}
\email{js8441@princeton.edu}
\affiliation{Department of Chemistry, Princeton University, Princeton, NJ USA}

\date{\today}

\begin{abstract}
	We present a phase-space electronic Hamiltonian $\hH_{PS}$ (parameterized by both nuclear position $\bX$ and  momentum $\bP$) that  boosts each electron into the moving frame of the nuclei that are closest in real space -- without presuming the existence of an atomic orbital basis. We show that $(i)$ quantum-classical dynamics along such a Hamiltonian maintains momentum conservation and $(ii)$ diagonalizing such a Hamiltonian can recover the electronic momentum and electronic current density reasonably well.  In conjunction with other reports in the literature that such a phase-space approach can also recover vibrational circular dichroism (VCD) spectra, we submit that the present phase-space approach offers a testable and powerful approach to post-Born-Oppenheimer electronic structure theory.  Moreover, the approach is inexpensive and can be immediately applied to simulations of chiral induced spin selectivity experiments (where the transfer of angular momentum between nuclei and electrons is considered critical).
\end{abstract}

\maketitle 
 
\section{Introduction}
Linear and angular momentum transfer between nuclear and electronic degrees of freedom underlies a wide range of nonadiabatic or non-Born-Oppenheimer phenomena, including paramagnetic spin lattice relaxation, vibrational circular dichroism, and perhaps chiral-induced spin-selectivity (CISS) \cite{naaman:2012:jpcl,  fransson:2020prb:vibrational, gersten_nitzan:2013:jcp_chiral}.
 Indeed, in the context of recent CISS experiments by the Wasielewski group\cite{wasielewski:2023:science:ciss} in solution (i.e. not near a metal), it has become clear that there must be a non-trivial exchange of angular momentum between electronic spins and nuclear motion in some photoexcited processes. As such, there is at present an enormous push to develop nonadiabatic methods that can accurately incorporate such non-Born-Oppenheimer (BO) momentum transfer between electrons and nuclei.\cite{gross:2022:prl:angmom}

Now, when studying  momentum transfer between electrons and nuclei, especially with a focus on spin processes where angular momentum comes in small chunks of $\hbar$, one cannot emphasize enough that proper simulation methods must conserve the total momentum. Ironically, however, almost all modern quantum-classical dynamics fail to enforce such conservation\cite{truhlar:2020:project_rot_nac,coraline:2024:jcp:ehrenfest_conserve}.  The problem at bottom is that, within the BO approximation,  electrons are treated at a very different level of theory from nuclei; not only are the nuclei posited to be slow, but they are also treated classically, and as a result, {\em BO simulations conserve nuclear momentum rather than the total  nuclear plus electronic momentum.}\cite{xuezhi:2023:total_ang_bomd}  As we pointed out recently, this conclusion is obvious if one considers the case of a radical molecule with an odd number of electrons in the presence of spin-orbit coupling, where standard BO  conserve the total nuclear angular momentum but not the electronic spin\cite{xuezhi:2023:total_ang_bomd}. 
In order to rectify BO's lack of momentum conservation, the usual prescription found in the literature is to include a  Berry pseudomagnetic force\cite{xuezhi:2023:total_ang_bomd}.  Indeed, over the last several years, several research groups in the chemical\cite{helgaker:2021:jcp:dynamics_in_magnetic_field,helgaker:2022:jcp_conservation_laws_magnetic_field} and physics communities \cite{vanderbilt:2024:berryforce} have explored {\em ab initio} BO dynamics that include Berry forces, in or out of magnetic fields, both in gas and solid phases. Unfortunately, however, running such {\em ab initio} dynamics has several limitations. ($i$) First, the cost of evaluating a Berry force can be quite large; ($ii$) Second, there is no feedback from the Berry force on the electronic dynamics, so that some electronic observables (like electronic momentum) remain uncorrected. ($iii$) Third and most importantly, there is no way to apply a  Berry force in a nonadiabatic context\cite{wu:jcp:2021:first_attempt_complex}, and thus these simulations cannot be used to study momentum transfer during passage through a curve crossings (which would represent a meaningful extension of Landau-Zener theory\cite{wittig:2005:landau_zener}).  Admittedly, one can in principle run Ehrenfest dynamics and, include a non-abelian force (as suggested by Takatsuka\cite{takatsuka:2005:jcp,takatsuka:2011:pccp_review_nonadiabatic} and Krishna\cite{krishna:2007:ehr_plus_berry}) so as to recover momentum conservation\cite{coraline:2024:jcp:ehrenfest_conserve}, but the cost of evaluating the non-abelian Berry curvature is enormous, and the introduction of mean-field dynamics introduces other approximations one would like to avoid (e.g., the incorrect equilibrium distribution\cite{tully:2005:detailedbalance,tully:2008:detailedbalance}). Exact factorization techniques\cite{abedi:2010:prl_exact_factorization,abedi:2012:jcp_exact_factorization,gross:2017:prx:born_oppenheimer_mass,agostini:2022:jpcl:abinitio} are a promising alternative avenue for future development where angular momentum is conserved\cite{gross:2022:prl:angmom}, but results have so far been limited to small systems.  
For all of these reasons, developing inexpensive fully momentum-conserving nonadiabatic equations of motion for nuclei and electrons is a priority for chemical and condensed matter physics.

\section{Phase Space Electronic Structure Methods}

In this vein, a very enticing alternative, inexpensive strategy to BO theory is to run dynamics along what we will call phase-space adiabats.  The notion of developing self-consistent approaches to electronic transitions through eikonal transformations goes back many decades\cite{micha:1983:self_consistent_eikonal,weinberg:1962:eikonal},
and similar approaches with time-dependent Hamiltonians were derived by Berry in the 1980's (labeled superadiabats\cite{berry:1987:superadiabat,berry:1990:superadiabat}),
but for our purposes, the cleanest notion of a phase space electronic Hamiltonian for quantum-classical dynamics was written down 15 years ago by Shenvi\cite{shenvi:2009:jcp_pssh}.
The basis idea of a phase-space electronic Hamiltonian is to diagonalize the electronic Hamiltonian that  depends parameterically on both nuclear position ($\bm{X}$) and momentum ($\bm{P}$).
Guided by the form of the Schrodinger equation in the usual BO basis (and the previous work or others\cite{berry:1987:superadiabat,berry:1990:superadiabat,micha:1983:self_consistent_eikonal,weinberg:1962:eikonal}), Shenvi diagonalized a phase-space Hamiltonian of the form:
\begin{widetext}
    \begin{align} 
    \hat{H}_{\rm Shenvi}(\bm{X},\bP) 
    = &\sum_{IJK,A} \frac{1}{2M_A} \left( \bm{P}_A \delta_{IJ} - i\hbar \bm{{d}}^A_{IJ} \right)\cdot
    \left( \bP_A \delta_{JK} - i \hbar \bm{{d}}^A_{JK} \right)\ket{\Phi_I}\bra{\Phi_K} 
     +\sum_K E_{KK}(\bm{X})\ket{\Phi_K}\bra{\Phi_K}, 
    \label{eq:shenvi} 
\end{align}
\end{widetext}

A brief interlude about notation: Here, we let  $A,B,\cdots$ denote the index of atoms, $\bm{P}_A$ denotes the nuclear momentum of atom $A$, and more generally, all quantities that are {\bf bold} represent vectors (or matrix when stated explicitly) in ${\mathbb R}^3$ Euclidean space. $\bm{X}$ is  shorthand for $\{\bm{X}\} = (\bm{X}_A,\bm{X}_B,\cdots)$ which represents the collection of all nuclear positions and $\bm{P}$ is  shorthand for $\{\bm{P}\} = (\bm{P}_A,\bm{P}_B,\cdots)$ which represents the collection of all nuclear momenta. $I,J,K$ index electronic states. $\bm d^{A}_{JK} = \bra{\Phi_J}\pp{}{\bm{X}_A}\ket{\Phi_K}$ is the derivative coupling vector between the eigenstates $\ket{\Phi_J}$ and $\ket{\Phi_K}$ of the conventional electronic Hamiltonian, where $\hat{H}_{el}\ket{\Phi_K}=E_{K}\ket{\Phi_K}$. Although they do not appear in  Eq.~\ref{eq:shenvi},  $\hbm{x}$, $\hbm{p}$, $\hbm{l}$, and $\hbm{s}$ will denote the electronic position, linear momentum, orbital angular momentum, and spin operators, respectively. All quantities with a hat are electronic operators.  

Trajectories propagated along the eigensurfaces from diagonalizating the Shenvi Hamiltonian in Eq. \ref{eq:shenvi} do conserve total momentum.\cite{yanzewu:2024:jcp:pssh_conserve} Unfortunately, however, these dynamics also suffer from several problems.  First, the derivative coupling vector is known to diverge near avoided crossings, leading to numerical instability doing dynamics using the phase-space Hamiltonian in Eq.\ref{eq:shenvi}\cite{izmaylov:2016:jpc_dboc_pssh}. Second, such derivative couplings are not even well-defined in the case of degnerate states. Third, building the phase-space Hamiltonian in Eq. \ref{eq:shenvi} involves computing $N(N+1)/2$ derivative coupling vectors, each of which can be computationally expensive, and for dynamics one would require the {\em derivatives} of these vectors -- which would be even more expensive. 
Motivated by these problems, in Ref. \citenum{coraline:2024:jcp:pssh_conserve}, we demonstrated that, in an atomic orbital (AO) basis, one can construct many other phase-space Hamiltonians (where we approximate the derivative couplings $\bd_{IJ}^A$ in Eq. \ref{eq:shenvi} by a to-be determined operator $\hbm{\Gamma}^A$) that can satisfy momentum conservation provided the Hamiltonian is solved in an AO basis.   In this letter, we will now show that momentum conservation requires no such AO basis, and we will demonstrate that a meaningful Hamiltonian can indeed be constructed by boosting each electron into a frame of the local nucleus.

If $\hH_{el}$ is the usual electronic Hamiltonian,  consider a phase-space electronic Hamiltonian of the form
\begin{widetext}
\begin{align}
\hH_{\rm PS}(\bm{X},\bP) &= \sum_{A}\frac{1}{2M_A} \left( \bm{P}_A  - i\hbar \hbm{\Gamma}_A(\bX) \right)\cdot
    \left( \bP_A - i \hbar \hbm{\Gamma}_A(\bX) \right)+  \hH_{el}(\bm{X}),\label{eq:PSH} 
\end{align}
\end{widetext}
with eigenvalues and eigenvectors
 \begin{align}   
&\hat{H}_{\rm PS}(\bm{X},\bP)  \ket{\Psi_{\rm PS}(\bm{X},\bP)} = E_{\rm PS}(\bm{X},\bP) \ket{\Psi_{\rm PS}(\bm{X},\bP)}, \label{eq:ps_eig}
\end{align}
The equations of motion for the corresponding Hamiltonian mechanics are:
\begin{align}
\label{eq:hamilton}
\dot{\bX} = \frac{\partial E_{\rm PS}}{\partial \bP} \; \; \; \; \; 
\dot{\bP} = -\frac{\partial E_{\rm PS}}{\partial \bX}.
\end{align}
Obviously, such Hamilton's equations conserve energy. In what follows, we will prove that the dynamics in Eq. \ref{eq:hamilton} 
conserve linear and angular momentum  provided that the
 $ \hbm{\Gamma}_A$ operator satisfies four constraints:
 \begin{align}
    -i\hbar\sum_{A}\hbm{\Gamma}_{A} + \hbm{p} &= 0,\label{eq:Gamma1}  \\
    \Big[-i\hbar\sum_{B}\pp{}{\bm{X}_B} + \hbm{p}, \hbm{\Gamma}_A\Big] &= 0,\label{eq:Gamma2}\\
    -i\hbar\sum_{A}{\bm X}_{A} \times \hat{\bm \Gamma}_{A} + \hbm{l} + \hbm{s} &= 0,\label{eq:Gamma3}\\
     \Big[-i\hbar\sum_{B}\left(\bm{X}_B \times\pp{}{\bm{X}_B}\right)_{\gamma} + \hat{l}_{\gamma} + \hat{s}_{\gamma}, \hat{\Gamma}_{A \delta}\Big] \nonumber&\\
     = i\hbar \sum_{\alpha} \epsilon_{\alpha \gamma \delta} \hat{\Gamma}_{A \alpha},\label{eq:Gamma4}
\end{align} 
where ``$\times$'' represents the cross product between two vectors in ${\mathbb R}^3$. 
Note that Eqs. \ref{eq:Gamma1}-\ref{eq:Gamma4} are satisfied exactly by the derivative couplings (if we substitute $\bd^{A}$ for $\bGamma$ as in Eq. \ref{eq:shenvi})\cite{littlejohn:2023:jcp:angmom}. 
Intuitively, Eqs. \ref{eq:Gamma1} and \ref{eq:Gamma3} insist that the electronic eigenfunctions of $\hH_{\rm PS}$ will be expressed in coordinates relative to the nuclei: when the nuclei translate, the electrons gain momentum (Eq. \ref{eq:Gamma1});  when the nuclei rotate, the electrons gain angular momentum (Eq. \ref{eq:Gamma3}).
Eqs. \ref{eq:Gamma2} and \ref{eq:Gamma4} dictate that the form of the $\hbm{\Gamma}_A$ operators are unchanged if we displace or reorient the molecule (and see SM\cite{SM} sec.II for a quick proof). 

To prove momentum conservation, we evaluate the total linear and angular momentum (which are the sums of  the nuclear plus electronic components):
\begin{align}
    \bm P^{\rm tot} =&    \sum_{A} M_{A} \dot{\bm X}_{A} + \bra{\Psi_{\rm PS}(\bm{X},\bP)}\hbm{p} \ket{\Psi_{\rm PS}(\bm{X},\bP)}, \label{eq:Ptot}\\
    \bm L^{\rm tot} =&\sum_{A} M_{A} {\bm X}_{A}\times\dot{\bm X}_{A}\nonumber
    \\ &+ \bra{\Psi_{\rm PS}(\bm{X},\bP)}\hbm{l} +\hbm{s}  \ket{\Psi_{\rm PS}(\bm{X},\bP)}.\label{eq:Ltot}
\end{align}
Using Eqs. \ref{eq:Gamma1}, it follows that the nuclear kinetic momentum is 
\begin{equation}
\label{eq:kinetic}
    M_{A} \dot{\bm X}_{A} = \bm P_{A} - i\hbar \bra{\Psi_{\rm PS}(\bm{X},\bP)}\hbm{\Gamma}_{A} \ket{\Psi_{\rm PS}(\bm{X},\bP)}
\end{equation}
If one plugs  Eqs. \ref{eq:kinetic}, \ref{eq:Gamma1} and \ref{eq:Gamma3} into Eqs. \ref{eq:Ptot} and \ref{eq:Ltot}, the result is $\bm P^{\rm tot} =  \sum_{A} \bm P_{A} $ and $\bm L^{\rm tot} =   \sum_{A}  \bm L_{A} $.  Finally, momentum conservation follows from Eq. \ref{eq:hamilton} by noting that 
the phase-space energy $E_{\rm PS}(\bm{X},\bP)$ is invariant to translation and rotation (see SM\cite{SM} sec.II).


At this stage, the only remaining questions are ($i$) how to properly choose the $\hbm{\Gamma}_A$ operators in Eq. \ref{eq:PSH} and ($ii$) how to judge the merit of that choice.  We will address question ($i$) shortly, but as for question ($ii$), we are aware of two means for evaluating the accuracy of a phase-space Hamiltonian and its advantages over a BO electronic Hamiltonian. First, as is well known\cite{nafie:1983:jcp:el_momentum,patchkovskii:2012:jcp:electronic_current}, one can show that if one properly includes non-Born-Oppenheimer effects to first order, an electronic eigenstate ($\ket{\Phi}$) with zero momentum ($\left< \Phi \middle| \hbm{p} \middle| \Phi \right> = \bm{0}$) is perturbed into an electronic eigenstate ($\ket{\Psi}$) with non-zero momentum equal to: $\braket{\hat{\bm{p}}} =md\langle\hat{\bm x}\rangle/dt$.
Second, a strong phase-space electronic approach should yield a good estimate for the electronic current density, which is essential for recovering the missing velocity-form of the electronic dipole moment and the magnetic dipole moment within BO theory.\cite{nafie:1983:jcp:el_momentum,takatsuka:2021:jcp:flux_conservation} These quantities contribute directly to  the electronic component of the rotatory strength, that is observed in vibrational circular dichroism (VCD)\cite{nafie:1992:vcd};
These two constraints will allow us to verify the validity of any phase space Hamiltonian. In the present letter, we will investigate how well one recovers electronic momentum and electronic current density according to the continuity equation (shown in SM\cite{SM} sec.IV); { in a separate paper, we demonstrate that accurate VCD signals can indeed be recovered with this phase space approach \cite{duston:2024:jctc_vcd,Tao:2024:jcp_vcd}. 
}

\section{Form of $\bGamma$ in Eq. \ref{eq:PSH}}
The simplest physical model for the $\bGamma$-couplings in Eq. \ref{eq:PSH} arises from the realization that in an AO basis, whenever a nucleus moves, that nuclear motion drags along an electronic orbital that leads to a trivial derivative coupling.  As an example, consider a hydrogen atom moving in free space in the $x-$direction with an electron occupying a hydrogenic $3p_x$ orbital that translates with the nucleus. If one evaluates the derivative coupling between the hydrogenic $3p_x$ and $1s$ orbitals, one finds a nonzero value -- even though such a value is clearly eliminated by moving to a center of mass frame. This trivial component of the derivative coupling was recognized long ago and is often labeled a ``Bates factor''\cite{bates:1958:etf} or  
``electron translation factor'' (ETFs)\cite{delos:1978:pra:etf,errea:1994:etf,ohrn:1994:rmp:etf,winter:1982:pra:etf,schneiderman:1969:pr:etf,Fatehi:2011:ETF}.
Moreover, although translations are much easier to evaluate than rotations, the same physical mechanism also occurs whenever the nuclei of a given molecule rotate, and one can also identify  ``electron rotation factors'' (ERFs)\cite{athavale:2023:erf,tian:2024:jcp:erf} as well.  

Thus, a natural choice for the $\bGamma$-couplings is to set
\begin{align}
	\label{eq:easy}
    \hbm{\Gamma}_A = \sum_i \bm{\Gamma}_A'(\bm{\hat x}_i,\bm{\hat p}_i) + \bm{\Gamma}_A''(\bm{\hat x}_i,\bm{\hat p}_i),
\end{align}
where the sum over $i$ represents a sum over electrons, $\hbm{\Gamma}_A'$ is an ETF operator and $\hbm{\Gamma}_A''$ is an ERF operator.  Encouragingly, within the context of quantum chemistry calculations in an AO basis (where explicit forms for ETFs and ERFs are known), we have recently demonstrated that phase-space dynamics using Eq. \ref{eq:easy} recovers the correct electronic momentum\cite{coraline:2024:jcp:pssh_conserve} as well as quite accurate VCD signals for three small chiral molecules\cite{duston:2024:jctc_vcd}.
Unfortunately, however, all previous phase-space calculations relied on an 
AO basis (in order to construct $\hbm{\Gamma}$) and thus could not probe solid state systems in a plane-wave basis (which is very limiting). Furthermore, empirically one finds that using the $\hbm{\Gamma}$ from Ref. \citenum{tian:2024:jcp:erf}, it is difficult to converge VCD results as a function of basis set size (which is both a theoretical and practical limitation)\cite{duston:2024:jctc_vcd}. 

To move beyond an AO basis, let us make the ansatz that we divide up all of free space according to regions labeled by the closest nucleus (in the spirit of a Voronoi diagram\cite{Senechal:1993:voronoi}), and electrons are boosted relative to the motion of the nucleus that labels their location.
Mathematically, we imagine constructing a space-partitioning operator  
 $\Theta_A(\hbm{x})$ ($\hat{\Theta}_A$ as a shorthand) that satisfies:
\begin{itemize}
    \item $\hat{\Theta}_A \equiv \Theta_A(\hbm{x})$ is positive definite.
    \item $\sum_A \Theta_A(\hbm{x}) = 1$.
    \item $\Theta_A(\bm{x}) \gg \Theta_B(\bm{x})$ for $||\bm{x}-\bm{X}_A|| \ll ||\bm{x}-\bm{X}_B||$.
\end{itemize}
One example of this kind is to define $\Theta_A(\hbm{x})$ as
\begin{align}
\label{eq:theta}
    \Theta_A(\hbm{x}) = \frac{w_Ae^{-|\hbm{x}-\bm{X}_A|^2/\sigma_A^2}}{\sum_B w_Be^{-|\hbm{x}-\bm{X}_B|^2/\sigma_B^2}}.
\end{align}
where $w_A$ and $\sigma_A$ are parameters to be determined.
As derived in Sec. I of the SM\cite{SM}, ETF and ERF operators satisfying Eqs. \ref{eq:Gamma1}-\ref{eq:Gamma4} can then be defined as follows:
\begin{align}
    \hbm{\Gamma}_A' &= \frac{1}{2i\hbar}\left( \Theta_A(\hbm{x})\hbm{p}+\hbm{p}\Theta_A(\hbm{x})\right)\label{eq:etf_final},\\
    \hbm{\Gamma}_A'' &= -\sum_B\zeta_{AB}\left(\bm{X}_A -\bm{X}^0_{B}\right)\times \left(\bm{K}_B^{-1}\hbm{J}_B\right),\label{eq:erf_final}
\end{align}
where
\begin{widetext}
\begin{align}
    \hbm{J}_B &= \frac{1}{2i\hbar}\left((\hbm{x}-\bm{X}_B)\times\Theta_B(\hbm{x})\hbm{p}+(\hbm{x}-\bm{X}_B)\times\hbm{p}\Theta_B(\hbm{x})+2\hbm{s}\hat{\Theta}_B\right)\\
    \bm{X}_{B}^0 &=\frac{\sum_A\zeta_{AB}\bm{X}_A}{\sum_A\zeta_{AB}},\\
    \label{eq:KB}
    \bm{K}_B &=\sum_A\zeta_{AB}\left((\bm{X}_A^\top\bm{X}_A-\bm{X}_B^{0\top}\bm{X}_B^0)\mathcal{I}_3 - (\bm{X}_A\bm{X}_A^\top-\bm{X}_B^0\bm{X}_B^{0\top})\right).
\end{align}   
\end{widetext}
Here the function $\zeta_{AB} = \zeta(\bX_A,\bX_B)$ is a function of $|\bX_A-\bX_B|$  (though not necessarily a symmetric function of $A$ and $B$). 
In practice, for $\hbm{\Gamma}_A''$ to operate locally on electrons near atom A, we require $\zeta_{AB}$ to vanish for large separations of   $\bX_A$ and $\bX_B$, and we set: 
\begin{align}
\label{eq:zeta}
    \zeta_{AB} & = {\kappa_A} e^{-|\bm{X}_A-\bm{X}_B|^2/\beta_{AB}^2},
\end{align}
All that remains is to fix the parameters $w_A$ and $\sigma_A$ (in Eq. \ref{eq:theta}) and {$\kappa_A$ and} $\beta_{AB}$ (in Eq. \ref{eq:zeta}).  
For the $w_A$ {and $\kappa_A$} parameters in Eqs. \ref{eq:theta} {and Eq. \ref{eq:zeta}}, it is natural to weight by mass and set {$w_A = \kappa_A = M_A$.} {Note that, if $\zeta_{AB} = M_A$, $\bX_{B}^0$ is the center of mass and $\bK_{B}$ is the total nuclear moment of inertia matrix relative to the center of mass; }see discussion below.
 The most important parameter is  $\sigma_A$ (which incorporates the electronic extent of each atom). In principle, one would like to choose this length to be  $r_A^{vdw}$  (where $r_A^{vdw}$ is the van der Waals radius of atom $A$) but, as we will show below, one must be careful in this choice. A better (and more revealing) approach is to scale the radius of each atom by a factor $\lambda$, $\sigma_A \equiv \lambda r^{vdw}_{A}$, which will teach us a great deal about how phase space methods boost electrons around the motion of  the closest nuclei.
 For $\beta_{AB}$, we choose  $\beta_{AB} = \sqrt{2} \left(\sigma_A+\sigma_B\right)$.

\section{Results}
\begin{figure}
\includegraphics[width=0.48\textwidth]{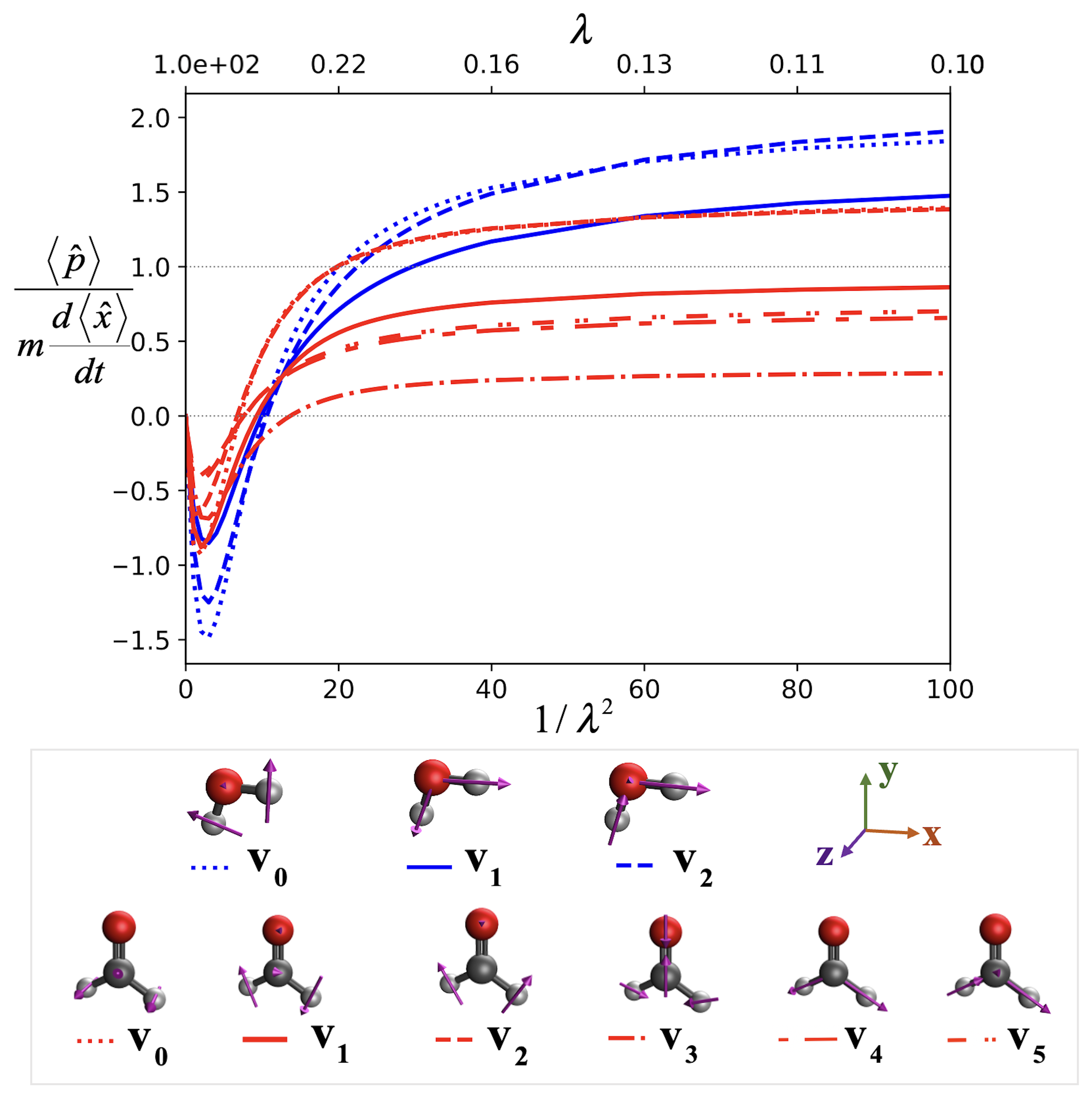}
\caption{  A plot of the ratio of the dominant Cartesian component of the  electronic momentum $\langle\hat{\bm p}\rangle$ with respect to the finite difference benchmark. The results of the water and formaldehyde are shown in blue and red, respectively. In principle, if one were to use $\bGamma = \bd$ as in Eq. \ref{eq:shenvi}, one would find the standard Ehrenfest result  
($ \langle\hat{\bm p}\rangle = m d \langle\hat{\bm x}\rangle/dt $), and thus this graph highlights that a simple phase space approach can recover a strong estimate of the electronic momentum (with the correct sign) using only an approximate $\bGamma$.} \label{fig:lambda}
 \end{figure}

The power of a phase-space approach (and the $\hbm{\Gamma}$ operator defined in Eqs. \ref{eq:etf_final}-\ref{eq:KB}) is highlighted in Fig. \ref{fig:lambda}, where we plot the electronic momentum $\langle\hat{\bm p}\rangle$ for the vibration modes of  small organic molecules and benchmarked against the finite-difference calculations 
$md\langle\hat{\bm x}\rangle/dt$. For these calculations, we probe the effect of making each atomic radius 
smaller ($1/\lambda^{2} \rightarrow 0$) 
or larger ($1/\lambda^{2}\rightarrow 100$).  
As shown in Fig. \ref{fig:lambda}, we sampled the three vibrational modes of water as well as the six vibrational modes of formaldehyde. 
The electronic momentum calculated with the phase space approach in Fig. \ref{fig:lambda} includes both the ETFs ($\hbm{\Gamma}'$) and ERFs ($\hbm{\Gamma}''$) contributions. 
In practice, we find that the effect of the ERFs ($\hbm{\Gamma}''$) is more important for bending modes than for the stretching modes. 

Two conclusions stand out from Fig. \ref{fig:lambda}. First, note that, as $1/\lambda^{2} \rightarrow 0$, the electronic momentum becomes small. This feature follows by construction from our choice $w_A = \kappa_A = M_A$ (in Eqs. \ref{eq:theta} and \ref{eq:zeta}) because, in this limit, $1/\beta_{AB}^2 \rightarrow 0$ and all of the $\Theta_A$ terms become $M_A/M_{\rm tot}$.   {In such a case, it follows that  $-i\hbar\sum_A \frac{ \bm P_A\cdot\hbm{\Gamma}_A}{M_A} =  -\frac{\sum_A \bm P_A}{M_{\rm tot}}\cdot \bm{\hat p} - \left(\sum_A (\bX_A-\bX_{\rm CM}) \times \bP_A\right) \cdot \bI_{\rm CM}^{-1} \cdot \left( \hbm{l}_{\rm CM} +\hbm{s} \right)$. Here, $\bX_{\rm CM}$ is the center of mass, $\bI_{\rm CM}^{-1}$ is the total nuclear moment of inertia matrix relative to the center of mass, and $\hbm{l}_{\rm CM}$ is the electronic angular momentum relative to the center of mass.  Thus, in this limit, the phase space Hamiltonian couples the total electron linear momentum to the total nuclear linear momentum and  the total electronic angular momentum to the total nuclear angular momentum (all relative to the center of mass).} Because  normal modes do not involve a change in center of mass location or overall orientation, we conclude that (in this limit) we should predict zero electronic momentum along a normal mode (and indeed we do).
Second, in the opposite limit  (large $\lambda$), the values of the electronic momentum become relatively constant (and always have the correct sign), which is extremely encouraging. {Note that, if $\lambda$ gets too small, numerical convergence can become difficult unless one uses a tight grid; at present, we evaluate $\Theta$  on a Lebedev grid designed to sample density near that atomic nucleus for DFT calculations, but there is no reason we could not build a new grid to best sample $\Theta_A$.  Thus,  as long as  the partitioning is not too flat, and provided we have enough grid points to model the switching region,} there is a huge range of parameters whereby the electronic momentum is both robust and meaningful. In practice, so far we find that optimal results are found if we set  $\lambda \le 0.2$ (so that for every atom $A$, $\sigma_A$ is not larger than $0.2r^{vdw}_A$).

Importantly, one should be not be very disheartened insofar as the fact that we do not recover exactly $ \langle\hat{\bm p}\rangle =m d/dt \, \langle\hat{\bm x}\rangle $ in Fig. \ref{fig:lambda} because, as shown in Ref. \citenum{coraline:2024:jcp:pssh_conserve}, this result follows exactly (analytically) if we set $\bGamma = \bd$ and sum over all excited states. Our goal in this letter is to generate  $\bGamma$ by approximating the derivative coupling, and here our assumption has been that the electronic density attempts to maintain its shape relative to the nuclei when the latter move (which requires the use of several parameters to be well-defined). We have not allowed for the fact that, when a nucleus moves, there could in fact be a massive redistribution of electronic density in a given adiabatic electronic state, e.g. an electron transfer that occurs when one moves through a 
curve crossing, which is the entire rationale for Marcus theory\cite{nitzanbook}.
The most striking conclusion from Fig. \ref{fig:lambda} is that, in the adiabatic limit (when adiabatic states are well separate energetically), this simple phase-space approach can capture so much of the electronic momentum. Moreover, in the sec. IV of the SM\cite{SM}, we show that our phase space electronic Hamiltonian can even recover the electronic current density (which is a resolution of the electronic momentum over three dimensional space).

\section{Discussion and Conclusions}

Our phase approach above is summed up by Eqs. \ref{eq:PSH}, \ref{eq:etf_final} and \ref{eq:erf_final}.
From a bird's eye view, this approach  can be viewed as a local strategy to remove the local center of mass from a messy quantum mechanical problem with many nuclei and electrons. 
For calculations of a small isolated molecule,  such a procedure is well-known and can be found in just about any quantum textbook: one transforms from a system of lab coordinates to a new set of coordinates where the electrons are expressed relative to the nuclear center of mass\cite{jensen,littlejohn:2023:jcp:angmom}. While this approach is quite logical for an isolated system, upon reflection this approach is suboptimal in the context of electronic structure theory in the condensed phase. Indeed, within electronic structure theory, one paradigm has always been that calculations of isolated systems should remain separate (which necessarily leads to size-consistency of the total energy)\cite{sherrillnotes2}. The approach above would seem to break such a paradigm because, for one molecule on earth and another on the moon, the standard separation would express all electronic coordinates relative to some point beyond the exosphere. By contrast, in some senses, the intuition of a phase space approach is that, in solving the electronic Schrodinger equation, one can effectively choose different electronic coordinates as relative to the closest nucleus. Put differently, every electron experiences the motion of the nuclei close by such that each composite (nuclear plus electronic) subsystem attempts to conserve the local momentum.

Looking forward, the data in Fig. \ref{fig:lambda} is extremely encouraging as an initial step towards a new and more powerful (but still very inexpensive) approach to 
electronic structure theory beyond
BO  theory.  As far as developing a practical method, clearly
the next step is to generate a large set of data, with many molecules at many different geometries, and then generate an optimal $\Theta$ partitioning of space to optimize the performance of the electronic momenta; if the $\Theta$ function in Eq. \ref{eq:theta} is accurate enough already, we will need only to optimize the radii of each atom. One can also develop parameters by comparing phase space results against  VCD spectra\cite{stephens:1985:jpcc_vcd,gross:2015:jcp:vcd_exact_factorization,Luber:2022:VCD}; {  note that the present method has already been validated against experimental VCD signals for (2S,3S)-oxirane-d2\cite{Tao:2024:jcp_vcd}. Finally, as one last piece of evidence, for a model problem where exact eigenstates can be calculated numerically, we have recently demonstrated that the present phase-space approaches can yield improved vibrational energies \cite{Bian:2024:BF_vib},} so that further benchmarking against highly accurate, perhaps rotationally refined, vibrational spectra may allow for further development of a meaningful set of parameters.

Armed with a meaningful phase space approach to electronic structure theory, one can imagine a wide swath of potentially exciting applications. As mentioned in the introduction, phase space dynamics conserve angular momentum, and one immediate application will be to investigate what can be learned by including spin-orbit coupling in our simulations and then running dynamics along phase space adiabats.  Indeed, in the context of a curve crossing, it is possible that one will recover new spin physics during an electron transfer event, when changes in the electronic orbital angular momentum must be balanced by changes in the nuclear and/or electronic spin. In particular, there is some hope that the present simulations will offer a new appproach towards understand the CISS effect.\cite{waldeck:2024:chemrev:ciss} 

Beyond coupled nuclear-electronic-spin simulations, another very promising avenue is to investigate superconductivity.  The underlying physics for many superconductive systems is electron-phonon coupling\cite{schriefferbook:superconductivity}.
Thus, it might be very interesting to investigate the behavior of a phase-space Hamiltonian that directly couples phonons to electrons for low-gap systems, especially solid-state or otherwise extended systems, where the role of electron-electron correlation is critical\cite{helgakerbook,furche:2017:arpc:rpa,mhg:2010:static_correlation_hextuples:jcp}.

\section{Acknowledgments}
We thank Titouan Duston and Mansi Bhati for helpful discussions. JES thanks Daniel Ezra Subotnik for a lifetime of probing questions. 

Zhen Tao and Tian Qiu contributed equally to this work.

\newpage
\begin{widetext}
\section{Appendix}
\subsection{Motivating the Form of the ETF Operators and ERF Operators}\label{sec-prove-satisfy}
In this section, we motivate the electron translation factor (ETF) and electron rotation factor (ERF) operators appearing in Eqs. 14 and 15 in the main body of the paper.

\subsection{ETF operators ($\hbm{\Gamma}'$)}
Given that, for linear momentum conservation, one requires that the ETF operators sum to the electronic momentum, i.e.:
\begin{align}
    \sum_A \hbm{\Gamma}_A' = \frac{1}{i\hbar}\hbm{p}\label{eq:etf_trans}
\end{align}

Let us multiply the R.H.S. of Eq. \ref{eq:etf_trans} on the left (and then on the right) by $\sum_A\hat{\Theta}_A$, and then take the average. The result is
\begin{align}
    \sum_A \hbm{\Gamma}_A'=\frac{1}{2i\hbar}\sum_A\left( \hat{\Theta}_A\hbm{p}+\hbm{p}\hat{\Theta}_A\right),
\end{align}
which suggests the following expression for $\hbm{\Gamma}_A'$:
\begin{align}
    \hbm{\Gamma}_A' = \frac{1}{2i\hbar}\left( \hat{\Theta}_A\hbm{p}+\hbm{p}\hat{\Theta}_A\right).\label{eq:etf}
\end{align}
It can be shown easily that $\hbm{\Gamma}_A'$ defined by Eq. \ref{eq:etf} is anti-hermitian (note that both $\hat{\Theta}_A$ and $\hbm{p}$ are hermitian), which is crucial for constructing a phase-space electronic Hamiltonian. 

\subsubsection{ERF operators ($\hbm{\Gamma}''$)}
Given the definition of $\bGamma'$ in Eq. \ref{eq:etf}, Eqs. 5 and 7 in the main body of the paper become
\begin{align}
    \sum_A \hbm{\Gamma}_A'' &= 0,\label{eq:erf_trans}\\
    \sum_A \bm{X}_A\times\hbm{\Gamma}_A'' &= \frac{1}{i\hbar}(\hbm{x}\times\hbm{p}+\hbm{s}) - \sum_A \bm{X}_A\times\hbm{\Gamma}_A'\nonumber\\
    &=\frac{1}{2i\hbar}\sum_A\left((\hbm{x}-\bm{X}_A)\times(\hat{\Theta}_A\hbm{p})+(\hbm{x}-\bm{X}_A)\times(\hbm{p}\hat{\Theta}_A)+2\hbm{s}\hat{\Theta}_A\right),\label{eq:erf_rot_0}
\end{align}
which are now constraints on $\hbm{\Gamma}_A''$. If one defines
\begin{align}
    \hbm{J}_A = \frac{1}{2i\hbar}\left((\hbm{x}-\bm{X}_A)\times(\hat{\Theta}_A\hbm{p})+(\hbm{x}-\bm{X}_A)\times(\hbm{p}\hat{\Theta}_A)+2\hbm{s}\hat{\Theta}_A\right),
\end{align}
Eq. \ref{eq:erf_rot_0} can be written succinctly as:
\begin{align}
    \sum_A \bm{X}_A\times\hbm{\Gamma}_A'' = \sum_A\hbm{J}_A.\label{eq:erf_rot}
\end{align}

Let us now show that if one defines $\hbm{\Gamma}_A''$ as
\begin{align}
    \hbm{\Gamma}_A'' = -\sum_B\zeta_{AB}\left(\bm{X}_A -\bm{X}^0_{B}\right)\times \left(\bm{K}_B^{-1}\hbm{J}_B\right),\label{eq:erf}
\end{align}
both Eqs. \ref{eq:erf_trans} and \ref{eq:erf_rot} can be satisfied given proper choice of the unknown vector $\bm{X}^0_{B}$ and a to-be-determined $3 \times 3$ matrix $\bm{K}_B$. In Eq. \ref{eq:erf}, in order to ensure that $\bGamma''_A$ is localized around atom $A$, we have introducting 
a weighting factor where $\zeta_{AB}$  that reaches {$M_A$}  when $A=B$ and decays rapidly otherwise. 

Inserting Eq. \ref{eq:erf} into Eq. \ref{eq:erf_trans}, one can rewrite the necessary constraint as:
\begin{align}
    \sum_{AB}\zeta_{AB}\left(\bm{X}_A -\bm{X}^0_{B}\right)\times \left(\bm{K}_B^{-1}\hbm{J}_B\right) = 0.
\end{align}
The simplest way to satisfy Eq. \ref{eq:erf_trans} is to set
\begin{align}
    \sum_A \zeta_{AB}\left(\bm{X}_A -\bm{X}^0_{B}\right) = 0,\label{eq:xb0}
\end{align}
which suggests that we define a local average position,
\begin{align}
    \bm{X}_{B}^0 =\frac{\sum_A\zeta_{AB}\bm{X}_A}{\sum_A\zeta_{AB}}.
\end{align}
Next, inserting Eq. \ref{eq:erf} into Eq. \ref{eq:erf_rot}, one finds the constraint has the form:
\begin{align}
    -\sum_A \bm{X}_A\times\left(\sum_B\zeta_{AB}\left(\bm{X}_A -\bm{X}^0_{B}\right)\times \left(\bm{K}_B^{-1}\hbm{J}_B\right)\right)=\sum_B\hbm{J}_B.
\end{align}
The most natural way to satisfy Eq. \ref{eq:erf_rot} is then to set
\begin{align}
    -\sum_A \bm{X}_A\times\left(\zeta_{AB}\left(\bm{X}_A -\bm{X}^0_{B}\right)\times \left(\bm{K}_B^{-1}\hbm{J}_B\right)\right) = \hbm{J}_B,
\end{align}
which suggests that we define the matrix $\bm{K}_B$ such that for any arbitrary vector $\bm{v}$, one has
\begin{align}
    \bm{K}_B\bm{v} &= -\sum_A \bm{X}_A\times\left(\zeta_{AB}\left(\bm{X}_A -\bm{X}^0_{B}\right)\times\bm{v}\right)\\
    \Rightarrow\bm{K}_B &= -\sum_A\zeta_{AB}\left( (\bm{X}_A -\bm{X}^0_{B})\bm{X}_A^\top - \bm{X}_A^\top(\bm{X}_A -\bm{X}^0_{B})\mathcal{I}_3\right)\\
    &=\sum_A\zeta_{AB}\left((\bm{X}_A^\top\bm{X}_A-\bm{X}_B^{0\top}\bm{X}_B^0)\mathcal{I}_3 - (\bm{X}_A\bm{X}_A^\top-\bm{X}_B^0\bm{X}_B^{0\top})\right),\label{eq:K_final}
\end{align}
where $\mathcal{I}_3$ is a 3 by 3 identity matrix. Note that we have invoked Eq. \ref{eq:xb0}  in deriving Eq. \ref{eq:K_final}.

Thus, we have now motivated our choice of functions Eqs. 14 and 15, and shown that these operators 
satisfy  Eqs. 5 and 7.

\subsection{A Demonstration that the $\bGamma'$ and $\bGamma''$ operators transform correctly under translations and rotations}\label{sec-prove-invariant}
\subsubsection{Proof of Eq. 6}
The proof of Eq. 6 is straightforward. Note that $\hbm{\Gamma}$ is a function of $ \hbm{x}_A \equiv \hbm{x}-\bm{X}_A$ and $ \bm{X}_{AB} \equiv \bm{X}_A-\bm{X}_B$ only, i.e.,
\begin{align}
    \hbm{\Gamma}_A = \hbm{\Gamma}_A\left(\{ \bm{x}_B\},\{\bm{X}_{BC}\}\right)
\end{align}

It follows that
\begin{align}
    \grad{\bm{x}}\hbm{\Gamma}_A &= \sum_B \grad{\bm{x}_B}\hbm{\Gamma}_A\\
    \sum_B\grad{\bm{X}_B}\hbm{\Gamma}_A &= -\sum_B \grad{\bm{x}_B}\hbm{\Gamma}_A
\end{align}
and therefore
\begin{align}
    \Big[-i\hbar\sum_{B}\pp{}{\bm{X}_B} + \hbm{p}, \hbm{\Gamma}_A\Big]
    =&-i\hbar\left(\sum_{B}\left(\grad{\bm{X}_B}\hbm{\Gamma}_A\right) + \left(\grad{\bm{x}}\hbm{\Gamma}_A\right)\right) \\
    =&0.
\end{align}

Another means to prove Eq. 6 is to notice that,
because the $\hbm{\Gamma}$ operators in Eqs. 14-15 are defined in terms of the electronic positions relative to the nuclei, 
\begin{align}
    \bGamma(\bm{x}+d\bm{s},\bm{X}+d\bm{s}) = \bGamma(\bm{x},\bm{X}),
\end{align}
it follows that the matrix representing $\bGamma$ must be the same  in any electronic basis of states that moves with the nuclear coordinates (e.g. the adiabatic basis\cite{littlejohn:2023:jcp:angmom,littlejohn:2024:jcp:moyal}). Thus, if $d\bm{s}$ denotes an infinitesimal translation of all atoms,  the following identity holds

\begin{align}
\braket{\Phi_I(\bm{X}+d\bm{s})|\hbm{\Gamma}_A(\bm{X}+d\bm{s})|\Phi_J(\bm{X}+d\bm{s})} - \braket{\Phi_I(\bm{X})|\hbm{\Gamma}_A(\bm{X})|\Phi_J(\bm{X})} = 0\label{eq:app:trans}
\end{align}
in any basis that satisfies
\begin{align}
\label{eq:stdphase_translate}
    \left(-i\hbar\sum_B\grad{\bm{X}_B}+\hbm{p}\right)\ket{\Phi_I(\bm{X})} =     -i\hbar\left(\sum_B\grad{\bm{X}_B}+\grad{\bm{x}}\right)\ket{\Phi_I(\bm{X})} = 0.
\end{align}
Eq. \ref{eq:stdphase_translate} is the standard phase relationship chosen for adiabats in Born-Oppenheimer theory\cite{littlejohn:2023:jcp:angmom,littlejohn:2024:jcp:moyal}.

Now, let us expand the 
 the left hands side (LHS) of Eq. \ref{eq:app:trans} to first order in $d\bm{s}$:
\begin{align}
LHS=&\left<\Phi_I(\bm{X})\middle|\sum_B\left(d\bm{s}\cdot\grad{\bm{X}_B}\hbm{\Gamma}_A(\bm{X})\right)\middle|\Phi_J(\bm{X})\right>\nonumber\\
&+\left<\sum_Bd\bm{s}\cdot\grad{\bm{X}_B}\Phi_I(\bm{X})\middle|\hbm{\Gamma}_A(\bm{X})\middle|\Phi_J(\bm{X})\right>+\left<\Phi_I(\bm{X})\middle|\hbm{\Gamma}_A(\bm{X})\middle|\sum_Bd\bm{s}\cdot\grad{\bm{X}_B}\Phi_J(\bm{X})\right>\\
=&\left<\Phi_I(\bm{X})\middle|\left[\sum_Bd\bm{s}\cdot\grad{\bm{X}_B},\hbm{\Gamma}_A(\bm{X})\right]\middle|\Phi_J(\bm{X})\right>\nonumber\\
    &+\frac{i}{\hbar}\left( \left<d\bm{s}\cdot\hbm{p}\Phi_I(\bm{X})\middle|\hbm{\Gamma}_A(\bm{X})\middle|\Phi_J(\bm{X})\right> - \left<\Phi_I(\bm{X})\middle|\hbm{\Gamma}_A(\bm{X})\middle|d\bm{s}\cdot\hbm{p}\Phi_J(\bm{X})\right> \right)\\
    =&\frac{i}{\hbar}\left<\Phi_I(\bm{X})\middle|\left[d\bm{s}\cdot\left(-i\hbar\sum_B\grad{\bm{X}_B}+\hbm{p}\right),\hbm{\Gamma}_A(\bm{X})\right]\middle|\Phi_J(\bm{X})\right>
    \label{eq:lastLHS_translate}
\end{align}

Because the translating basis of states $\left\{ \ket{\Phi_I} \right\}$ is assumed to be complete, the vanishing of the LHS in Eq. \ref{eq:lastLHS_translate} implies Eq. 6.

\subsubsection{Proof of Eq. 8}
 Eq. 8 can be proved in an analogous fashion. Let $\mathcal{R}(d\bm{\theta})$ be an infinitesimal rotation of $d\bm{\theta}$ and let us begin by showing that the $\bGamma$ operators in Eqs. 14-15 satisfy:
\begin{align}
    \bm{\Gamma}_A(\mathcal{R}\hbm{x},\mathcal{R}\hbm{p},\mathcal{R}\hbm{s},\mathcal{R}\bm{X}) = \mathcal{R}\bm{\Gamma}_A(\hbm{x},\hbm{p},\hbm{s},\bm{X})\label{eq:app:rot_equiv} 
\end{align}
To prove Eq. \ref{eq:app:rot_equiv}, the first two things to note is that the dot product between two vectors is invariant under $\mathcal{R}$, and the cross product between two vectors rotates with the vectors, i.e.,
\begin{align}
    \mathcal{R}\bm{v}_i\cdot\mathcal{R}\bm{v}_j &= \bm{v}_i\cdot\bm{v}_j\\
    \mathcal{R}\bm{v}_i\times\mathcal{R}\bm{v}_j &= \mathcal{R}(\bm{v}_i\times\bm{v}_j)
\end{align}
where $\bm{v}_i,\bm{v}_j$ are any vectors. It follows that
\begin{align}
    \bm{\Gamma}_A'(\mathcal{R}\hbm{x},\mathcal{R}\hbm{p},\mathcal{R}\bm{X}) &= \mathcal{R}\bm{\Gamma}_A'(\hbm{x},\hbm{p},\bm{X})\label{eq:app:rot_gamma1}\\
    \bm{J}_B(\mathcal{R}\hbm{x},\mathcal{R}\hbm{p},\mathcal{R}\hbm{s},\mathcal{R}\bm{X}) &=\mathcal{R}\bm{J}_B(\hbm{x},\hbm{p},\hbm{s},\bm{X})
\end{align}
Next, let $\hbm{V}_B = \bm{K}_B^{-1}(\bm{X})\hbm{J}_B$. A simple derivation shows that
\begin{align}
    \hbm{J}_B &= \bm{K}_B(\bm{X})\hbm{V}_B = \sum_A \bm{X}_A\times\left(\zeta_{AB}\left(\bm{X}_A -\bm{X}^0_{B}\right)\times\hbm{V}_B\right)\\
    \Rightarrow\mathcal{R}\hbm{J}_B &= \mathcal{R}\left(\sum_A \bm{X}_A\times\left(\zeta_{AB}\left(\bm{X}_A -\bm{X}^0_{B}\right)\times\hbm{V}_B\right)\right)\\
    &=\sum_A \mathcal{R}\bm{X}_A\times\left(\zeta_{AB}\left(\mathcal{R}\bm{X}_A -\mathcal{R}\bm{X}^0_{B}\right)\times\mathcal{R}\hbm{V}_B\right)\\
    &=\bm{K}_B(\mathcal{R}\bm{X})\mathcal{R}\hbm{V}_B\\
    \Rightarrow\mathcal{R}\hbm{V}_B &= \bm{K}_B^{-1}(\mathcal{R}\bm{X})\mathcal{R}\hbm{J}_B
\end{align}
from which one finds
\begin{align}
    \bm{\Gamma}_A''(\mathcal{R}\hbm{x},\mathcal{R}\hbm{p},\mathcal{R}\hbm{s},\mathcal{R}\bm{X}) &= \mathcal{R}\bm{\Gamma}_A''(\hbm{x},\hbm{p},\hbm{s},\bm{X})\label{eq:app:rot_gamma2}
\end{align}
Clearly, Eqs. \ref{eq:app:rot_gamma1} and \ref{eq:app:rot_gamma2} imply \ref{eq:app:rot_equiv}.

Eq. \ref{eq:app:rot_equiv} dictates that the matrix elements of the $\bGamma$ operators in Eqs. 14-15 are invariant to rotations in a basis that rotates with the system.  Mathematically, Eq. \ref{eq:app:rot_equiv} implies that

\begin{align}
    \braket{\Phi_I(\mathcal{R}(d\bm{\theta})\bm{X})|\hbm{\Gamma}_A(\mathcal{R}(d\bm{\theta})\bm{X})|\Phi_J(\mathcal{R}(d\bm{\theta})\bm{X})} = \braket{\Phi_I(\bm{X})|\mathcal{R}(d\bm{\theta})\hbm{\Gamma}_A(\bm{X})|\Phi_J(\bm{X})}\label{eq:app:rot}
\end{align}
in a basis that rotates with the molecule and has the standard phase convention:
\begin{align}
    \left(-i\hbar\sum_B\bm{X}_B\times\grad{\bm{X}_B}+\hbm{l}+\hbm{s}\right)\ket{\Phi_I(\bm{X})} = 0
\end{align}
Again, this phase convention is the standard choice of phase for the adiabatic states within the Born-Oppenheimer approximation\cite{littlejohn:2023:jcp:angmom,littlejohn:2024:jcp:moyal}.

At this point, let us expand Eq. \ref{eq:app:rot} in powers of $\theta$. Note that 
\begin{eqnarray}
    \mathcal{R}_{\alpha \beta}(d\bm{\theta}) =\delta_{\alpha \beta} + \sum_{\gamma} \epsilon_{\alpha \beta \gamma} d\theta_{\gamma}+ \mathcal{O}(d\bm{\theta}^2),
\end{eqnarray}
the Taylor series of the ket is:
\begin{align}
    \ket{\Phi_J(\mathcal{R}(d\bm{\theta})\bm{X})} - \ket{\Phi_J(\bm{X})} =&-\left(\sum_B\bm{X}_B\times\grad{\bm{X}_B}\right)\cdot d\bm{\theta}\ket{\Phi_J(\bm{X})} + \mathcal{O}(d\bm{\theta}^2)\\
    =&-\frac{1}{i\hbar}\left(\left(\hbm{l}+\hbm{s}\right)\cdot d\bm{\theta}\ket{\Phi_I(\bm{X})}\right)+ \mathcal{O}(d\bm{\theta}^2)
\end{align}
Subtracting $\braket{\Phi_I(\bm{X})|\hbm{\Gamma}_A(\bm{X})|\Phi_J(\bm{X})}$ from both side of Eq. \ref{eq:app:rot}, the L.H.S. gives (to the first order in $d\bm{\theta}$)
\begin{align}
    &\braket{\Phi_I(\mathcal{R}(d\bm{\theta})\bm{X})|\hbm{\Gamma}_A(\mathcal{R}(d\bm{\theta})\bm{X})|\Phi_J(\mathcal{R}(d\bm{\theta})\bm{X})} - \braket{\Phi_I(\bm{X})|\hbm{\Gamma}_A(\bm{X})|\Phi_J(\bm{X})}\\
    =&-\left<\Phi_I(\bm{X})\middle|\left[\left(\sum_B\bm{X}_B\times\grad{\bm{X}_B}\right)\cdot d\bm{\theta},\hbm{\Gamma}_A(\bm{X})\right]\middle|\Phi_J(\bm{X})\right>\nonumber\\
    &-\frac{i}{\hbar}\left(\braket{(\hbm{l}+\hbm{s})\cdot d\bm{\theta}\Phi_I(\bm{X})|\hbm{\Gamma}_A(\bm{X})|\Phi_J(\bm{X})} - \braket{\Phi_I(\bm{X})|\hbm{\Gamma}_A(\bm{X})|(\hbm{l}+\hbm{s})\cdot d\bm{\theta}\Phi_J(\bm{X})}\right)\\
    =&-\frac{i}{\hbar}\left<\Phi_I(\bm{X})\middle|\left[\left(-i\hbar\sum_B\bm{X}_B\times\grad{\bm{X}_B}+\hbm{l}+\hbm{s}\right)\cdot d\bm{\theta},\hbm{\Gamma}_A(\bm{X})\right]\middle|\Phi_J(\bm{X})\right>.
\end{align}
Regrading the R.H.S, it is easier to express it in explicit Cartesian indices:
\begin{align}
    &\left(\braket{\Phi_I(\bm{X})|\mathcal{R}(d\bm{\theta})\hbm{\Gamma}_A(\bm{X})|\Phi_J(\bm{X})} - \braket{\Phi_I(\bm{X})|\hbm{\Gamma}_A(\bm{X})|\Phi_J(\bm{X})}\right)_\delta\\
    =&\braket{\Phi_I(\bm{X})|\sum_{\beta\gamma}\epsilon_{\delta\beta\gamma}d\theta_\gamma\bm{\Gamma}_{A\beta}(\bm{X})|\Phi_J(\bm{X})}
\end{align}

Therefore, Eq. \ref{eq:app:rot} is equivalent to
\begin{align}
    &\left[\sum_{\gamma}\left(-i\hbar\sum_B\bm{X}_B\times\grad{\bm{X}_B}+\hbm{l}+\hbm{s}\right)_\gamma d\theta_\gamma,\hat{\Gamma}_{A\delta}(\bm{X})\right] = i\hbar\sum_{\alpha\gamma}\epsilon_{\delta\alpha\gamma}\hat{\Gamma}_{A\alpha}(\bm{X})d\theta_\gamma\\
    \Rightarrow&\left[\left(-i\hbar\sum_B\bm{X}_B\times\grad{\bm{X}_B}+\hbm{l}+\hbm{s}\right)_\gamma,\hat{\Gamma}_{A\delta}(\bm{X})\right]=i\hbar\sum_{\alpha}\epsilon_{\alpha\gamma\delta}\hat{\Gamma}_{A\alpha}(\bm{X})
\end{align}
which proves that Eqs. 14 and 15 satisfy Eq. 8.

\subsection{Computational details}
\label{sec-computation}
To generate the data in Fig. 1 in the main body of the paper,
the geometries of the water and the formaldehyde molecules are optimized at a standard restricted Hatree-Fock (RHF) level with a cc-pVTZ basis set. The vibrational modes in Fig. 1 are computed from the corresponding Hessian calculations. The benchmarks $md\braket{\hat{\bm x}}/dt $ are computed from finite-difference RHF calculations, and the linear momentum $\braket{\hat{\bm p}}$ are computed with RHF as well under the the phase-space framework, implemented in a developmental version of the Q-Chem electronic structure package.\cite{qchem6} The first-order and the second-order $\bm \Gamma$-couplings are evaluated on Lebedev grids. To ensure numerical accuracy, an unpruned (250, 974) grid is used to generate data in Fig. 1 so that constraints in Eqs. 5 and 7 are satisfied to less than $10^{-7} \hbar/a_{0}$ and $10^{-7} \hbar$, respectively in the atomic orbital basis. In all cases, we assume the nuclei are moving with momentum in directions corresponding to different vibrational modes and with magnitudes corresponding to the average momentum at temperature 100K. The finite-difference benchmark calculations use a time step of 0.0242 fs.

\subsection{Electronic Continuity Equation}
\label{sec-current}
Besides agreeing with VCD experiments and capturing a nonvanishing electronic momentum, we will now show that one of the essential validations of a phase space electronic Hamiltonian is the capacity to recover the missing continuity equation for the electronic motion.  Before showing results, let us review how the continuity equations for the current density can recovered using wavefunctions corrected for non-BO effects.\cite{nafie:1983:jcp:el_momentum,takatsuka:2021:jcp:flux_conservation} Thereafter, we will perform a simple numerical test for an \ce{H_{2}} molecule translating and stretching, from which we will observe that the continuity equation can reasonably satisfied using the $\bm \Gamma$ expression presented in Eqs. 14-15 -- though far worse results are recovered using a previously defined phase-space electronic Hamiltonian (from Ref. \cite{tian:2024:jcp:erf},\cite{coraline:2024:jcp:pssh_conserve}) based on an atomic orbital basis.

\subsubsection{Theory}
For a system of electrons and nuclei, let us begin by defining the electornic charge and the current density operators,
\begin{align}
        \hat{\rho}_{e}(\bm r) &= \hat{\psi}^{\dagger}(\bm r)\hat{\psi}(\bm r)\\
        \hat{\bm j}_{e}(\bm r) &= -\frac{i\hbar}{2m_{e}} \Big[\hat{\psi}^{\dagger}(\bm r)\nabla_{\bm r}\hat{\psi}(\bm r) - \Big(\nabla_{\bm r}\hat{\psi}^{\dagger}(\bm r)\Big)\hat{\psi}(\bm r)\Big]
\end{align}
where $\hat{\psi}^{\dagger}(\bm r)/\hat{\psi}(\bm r)$ is the creation and annihilation operator for an electron at position $\bm r$. As is well known, there is a continuity equation for the electronic based on the electronic Schrodinger's equation,
\begin{align}
\label{eq:drho_dt}
    \frac{d}{dt}\hat{\bm \rho}_e(\bm r)  = \frac{i}{\hbar}\left[ \bm \hat{H}_{e},\hat{\bm \rho}_e (\bm r)\right] = -\nabla \cdot \hat{\bm j}_e(\bm r),
\end{align}
Here we have written the electronic Hamiltonian operator in terms of the creation and annihilation operators,
\begin{align}
    \hat{H}_{e} = \int d^{3}\bm r \hat{\psi}^{\dagger}(\bm r) \Big[-\frac{\hbar^2}{2m}\nabla^{2}_{\bm r}+ V(\bm r)\Big]
    \hat{\psi}(\bm r) + \frac{1}{2}\int d^{3}\bm r \int d^{3}\bm r' \hat{\psi}^{\dagger}(\bm r) \hat{\psi}^{\dagger}(\bm r') \frac{q^2_{e}}{|\bm r-\bm r'|} \hat{\psi}(\bm r') \hat{\psi}(\bm r) 
\end{align}
Eq. \ref{eq:drho_dt} is an operator equation that holds for exact quantum dynamics.  One means of checking the accuracy of approximate dynamics, (e.g. for a trajectory moving along eigensurface of a phase space electronic Hamiltonian) is to test whether or not this continuity equation still holds during such approximate dynamics.
Eq. \ref{eq:drho_dt} clearly does not hold for BO motion because, in such a case, the electronic current operator will always have expectation value zero.

To that end, let  $\Phi$ be a BO eigenstate.  Treating the first-order non-BO operator as a perturbation and assuming non-degenerate states, one can construct a wavefunction corrected to the first order,
\begin{align}
\label{eq:psi}
    \ket{\tilde{\Psi}_I} = \ket{\Phi_I} - i\hbar \sum_{J\ne I} \frac{\bP}{\bm M}\cdot\frac{ \bracket{\Phi_J}{ \frac{\partial}{\partial \bm R} }{\Phi_I}}{E_I -E_J} \ket{\Phi_J}
\end{align}
Next, we can evaluate the expectation value of $\bra{\tilde{\Psi}_I}{\hat{\bm j}_e}(\bm r)\ket{\tilde{\Psi}_I}$,
\begin{align}
    \bra{\tilde{\Psi}_I}{\hat{\bm j}_{e}(\bm r)}\ket{\tilde{\Psi}_I}= 2 \hbar \text{Im}  \sum_{J\ne I} \frac{\bP}{\bm M}\cdot\frac{ \bracket{\Phi_J}{ \frac{\partial}{\partial \bm R} }{\Phi_I}}{E_I -E_J} \bra{\Phi_I}{\hat{\bm j}_{e}}(\bm r)\ket{\Phi_J}\label{eq:je_nad}
\end{align}
One notices immediately that, if we include the first-order non-BO correction, a non-zero current density arises.  

The next step is to evaluate the electronic flux for this current density.
From the commutation relationship in Eq.~\ref{eq:drho_dt}, it follows that 
\begin{align}
    (E_I - E_J) \bra{\Phi_I}{\hat{\bm \rho}_e(\bm r)}\ket{\Phi_J}= i\hbar   \bra{\Phi_I} \nabla \cdot{\hat{\bm j}_e(\bm r)}\ket {\Phi_J}\label{eq:rho_j}
\end{align}
Using Eq.~\ref{eq:rho_j}, we now evaluate 
\begin{align}
    \bra{\tilde{\Psi}_I}{\nabla \cdot \hat{\bm j}_{e}(\bm r)}\ket{\tilde{\Psi}_I}&= 2 \hbar \text{Im}  \sum_{J\ne I}\frac{\bP}{\bm M}\cdot\frac{ \bracket{\Phi_J}{ \frac{\partial}{\partial \bm R} }{\Phi_I}}{E_I -E_J} \bra{\Phi_I}{\nabla \cdot \hat{\bm j}_{e}}\ket{\Phi_J} \label{eq:jdiv_nad} \\
    & = -2 \text{Re}  \sum_{J } \frac{\bP}{\bm M}\cdot\frac{ \bracket{\Phi_J}{ \frac{\partial}{\partial \bm R} }{\Phi_I}}{E_I -E_J} \bracket{\Phi_I}{\hat{\bm \rho}_e(\bm r)}{\Phi_J} \label{eq:jdiv_nad2}\\
    & = - 2 \text{Re}  \bracket{\Phi_I}{ \hat{\bm \rho}_e (\bm r) \frac{  \bP}{\bm M} \cdot \frac{\partial}{\partial \bm R }}{\Phi_I} \label{eq:jdiv_nad3}\\
   \bra{\tilde{\Psi}_I}{ \nabla \cdot \hat{\bm j}_{e}(\bm r)}\ket{\tilde{\Psi}_I} & =  - \frac{d}{dt}  \bracket{\Phi_I} {\hat{\bm \rho}_e(\bm r)}{\Phi_I}\label{eq:j_rho}
\end{align}
Eq.~\ref{eq:j_rho} was a key result proven previously by Nafie\cite{nafie:1983:jcp:el_momentum} and  Takatuska.\cite{takatsuka:2021:jcp:flux_conservation}.  
According to Eq. \ref{eq:drho_dt}, in order to recover the electronic continuity equation, we need not to run exact quantum dynamics; we need only the first-order corrections to BO theory. We will now show that a phase space electronic Hamiltonian also recovers a meaningful electronic current density by plotting the left and right hand sides of Eq. ~\ref{eq:j_rho}.

As a side note, we mention that, from  Eq.~\ref{eq:j_rho}, one can derive the standard,  beyond-Born-Oppenheimer electronic momentum. In particular, we can multiply by position and electronic mass and integrate:
\begin{eqnarray}
m\int \bm r \nabla \cdot \bracket{\Phi_I} {\hat{\bm j}_{e}(\bm r)}{\Phi_I} d\bm r & = & -m \int  \bm r \frac{d}{dt}\bracket{\Phi_I} { \hat{\bm \rho}_e (\bm r)}{\Phi_I}d\bm r \label{eq:jandp1}\\
m\int  \bracket{\Phi_I} {\hat{\bm j}_{e}(\bm r)}{\Phi_I}  d\bm r & = &   m\frac{d}{dt}  \bracket{\Phi_I} {\int\bm r\hat{\psi}^{\dagger}(\bm r)\hat{\psi}(\bm r)d\bm r}{\Phi_I} \label{eq:jandp2}\\
\bracket{\Phi_I} {\hat{\bm p}_{e}(\bm r)}{\Phi_I} & = &  m \frac{d}{dt} \bracket{\Phi_I} {\hat{\bm r}}{\Phi_I} \label{eq:jandp3}
\end{eqnarray}

To go from the LHS of Eq. \ref{eq:jandp1} to the LHS of Eq. \ref{eq:jandp2}, we have integrated by parts. In Fig. 1 of the main paper, we benchmarked our phase space results against the result above (Eq. \ref{eq:jandp3}) for electronic momentum. In Figs. \ref{fig:h2_tran} and \ref{fig:h2_stretch} below, an even more strenuous test is to benchmark against the current density.

\subsubsection{Numerical Examples}
To test if and how a phase-space electronic Hamiltonian approach
obeys the continuity equation for the current density (Eq. \ref{eq:j_rho}), we first look at the translational motion of a \ce{H_{2}} molecule along the positive x-axis. In Fig. \ref{fig:h2_tran}, we plot  the difference in charge densities as computed by finite difference (left panel, the right hand side of Eq. \ref{eq:j_rho}) versus  the divergence of the current densities  (right panel, the left hand side of Eq. \ref{eq:j_rho}). We make this comparison for several different basis sets from small to large. Details of the basis sets used are listed in Table \ref{table:BSinfo}. 
The phase-space approach reproduce the correct nodal structures as we increase the size of the basis sets (in principle a complete set of basis is assumed in Eq. \ref{eq:drho_dt}), which is a good first validation of the method.
However, a stronger test of the method is to investigate an internal motion (rather a total translation).
\begin{table} [H]
    \centering
        \caption{Different Basis Sets for \ce{H_{2}}\label{table:BSinfo}}
        \begin{tabular} {|c|c|c|}
        \hline
         Basis Set & Num. and Type of Shell & \hfil  Num. of Basis Functions per H \\ 
        \hline
         STO-3G & 1s & 1   \\ 
        \hline
        \hfil cc-pVDZ (DZ)& 2s1p & 5 \\
                \hline
        \hfil cc-pVTZ(TZ) & 3s2p1d  &  14 \\
                        \hline
        \hfil aug-cc-pVTZ (aTZ)& 4s3p2d & 23 \\
                        \hline
        \hfil cc-pVQZ (QZ)&  4s3p2d1f &30 \\                   
        \hline
        \hfil aug-cc-pVQZ (aQZ)& 5s4p3d2f& 46 \\        
        \hline
        \end{tabular}
\end{table}
\begin{figure} [htbp]
\centering
\includegraphics[width=0.8\textwidth]{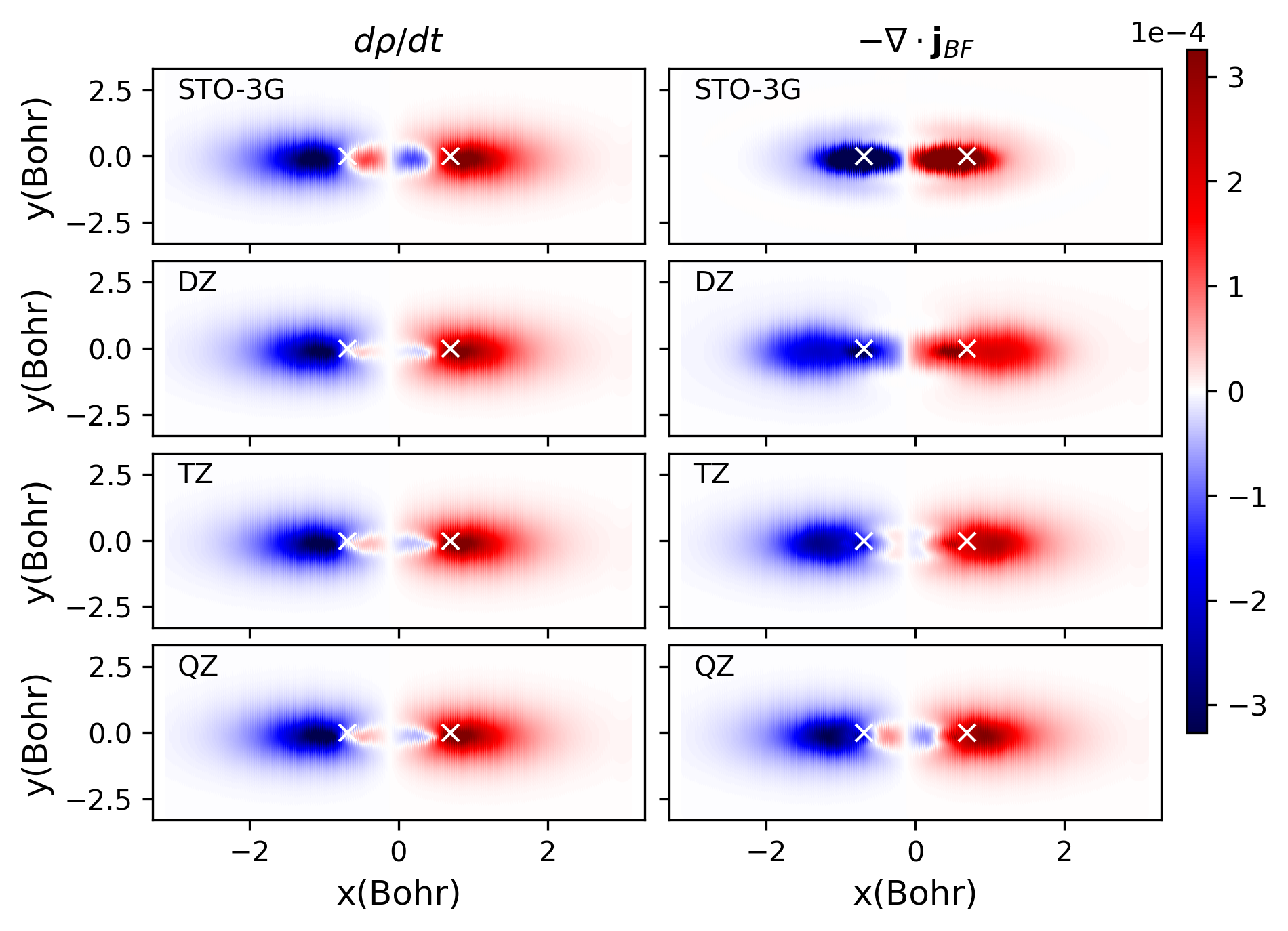} 
\caption{\label{fig:h2_tran} A test of the continuity equation for a translating \ce{H_{2}} molecule.  We plot the charge density differences (left panel) and the negative of the divergence of the current density  (with basis free (BF) $\hbm{\Gamma}$ from Eqs. 14-15, right panel) projected on the xy plane for different basis sets (small at the top, large at the bottom).  The positions of the two H atoms are at $\pm 0.69383$ Bohr along the x axis (indicated by the cross symbol in white). The strong agreement of these figures are a good validation of a phase-space electronic Hamiltonian.}
\end{figure}

To that end, next we look at results when we move one H atom closer to the other one, and we show how a poor electronic structure method can produce very low quality, non-physical current densities. In the spirit of the previous figure, on the left panel of Fig. \ref{fig:h2_stretch}, we plot the  differences in electronic charge densities computed by finite difference  and, in the middle panel, we plot the divergence of the current densities as computed from the $\hbm{\Gamma}$ operator presented in the main body of our paper (BF = basis free).  Lastly, on the right hand panel, we plot the divergence of the current density as found using the previous $\hat{\bm \Gamma}$  defined explicitly in atomic orbital basis (from Ref. \cite{tian:2024:jcp:erf} and \cite{coraline:2024:jcp:pssh_conserve}). Clearly, the left and middle panels do not depend strongly on basis; however, the right hand panel depends critically (and incorrectly) on basis and very bizarre nodal structures are found for the aQZ results in Fig. \ref{fig:h2_stretch}. Altogether, this data highlights the power of the method proposed in the main body of this manuscript, and also demonstrates why  a basis free approach for $\hat{\bm \Gamma}$ is essential when modeling coupled nuclear-electronic dynamics. 
For the basis-free $\hat{\bm \Gamma}$ results shown in Fig. \ref{fig:h2_stretch},  we use $1/\lambda^2=20$. 
As a side note, we note that a further tuning of the parameters (e.g. $1/\lambda^2=5$) can even remove the small extra nodal structures in the divergences of the current densities when computed with basis-free $\hat{\bm \Gamma}$. A more careful parameterization of the $\Theta$ function in Eq. 13 for a diverse group of molecules will be very fruitful for future work.

\begin{figure} [p]
\includegraphics[width=0.9\textwidth]{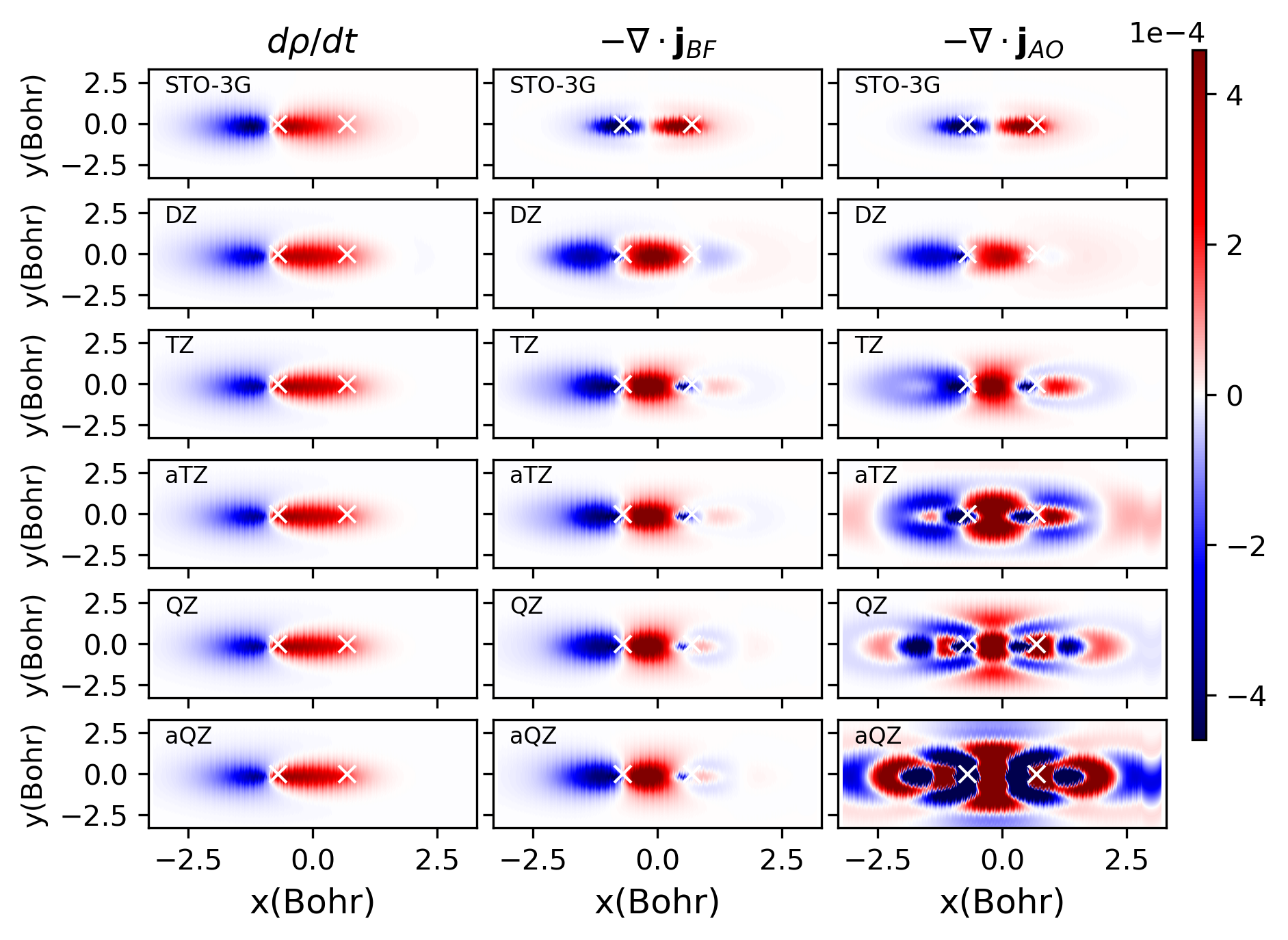}
\centering
\caption{\label{fig:h2_stretch} A test of the continuity equation for a stretching \ce{H_{2}} molecule where we move the H atom located at the negative x-axis closer to the other H atom with a velocity corresponding to the room temperature. We plot the charge density differences (left panel) and the negative of the divergence of the current density (right panel) projected on the xy plane as computed with different basis sets (small at the top, large at the bottom).   The positions of the two H atoms are at $\pm 0.69383$ Bohr along the x axis (indicated by the cross symbol in white). The agreement between the left and middle panels (with basis free (BF)  $\hbm{\Gamma}$ from Eqs. 14-15) offers a resounding validation of the approach taken in this paper. Note that even better results can be found by tuning $\lambda$ in Eq. 13. By contrast, one finds much worse results using the  phase-space electronic Hamiltonian approach using a  $\hbm{\Gamma}$ operator built from an (AO) atomic orbital basis (from Ref. \cite{tian:2024:jcp:erf} and \cite{coraline:2024:jcp:pssh_conserve}, right panel); the latter can clearly predict quite unphysical results for large basis sets.  }
\end{figure}

\end{widetext}
 \clearpage
\bibliography{ref}

\end{document}